\documentclass[12pt]{iopart}

\usepackage{iopams}
\usepackage[dvips]{graphicx}% Include figure files
\begin{document}

\title{Incommensurate Spin Density Wave versus local magnetic inhomogeneities in Ba(Fe$_{1-x}$Ni$_x$)$_2$As$_2$: a $^{57}$Fe M\"ossbauer spectral study}
\author{A Olariu, P Bonville, F Rullier-Albenque, D Colson and A Forget}
\address{Service de Physique de l'Etat Condens\'e, Orme des Merisiers, IRAMIS, CEA-Saclay (CNRS URA 2464), 91191 Gif sur Yvette Cedex, France}
\ead{pierre.bonville@cea.fr}

\begin{abstract}
We report $^{57}$Fe M\"ossbauer spectral results in pure and doped Ba(Fe$_{1-x}$Ni$_x$)$_2$As$_2$ with $x=0.01$ and 0.03. We show that all these materials present a first-order magnetic transition towards a magnetically ordered state. In the doped compounds, a broad distribution of Fe hyperfine fields is present in the magnetic phase. We successfully fit the M\"ossbauer data in Ba(Fe$_{1-x}$Ni$_x$)$_2$As$_2$ in the framework of two different models: 1) an incommensurate spin density wave; 2) a dopant-induced perturbation of the Fe polarization, recently proposed to interpret $^{75}$As NMR data in Ba(Fe$_{1-x}$Ni$_x$)$_2$As$_2$, which is valid only in the very dilute limit $x=0.01$. Moreover, we show here that these NMR data can also be successfully analysed in terms of the ``incommensurate model'' for all doping contents by using the parameters obtained from the M\"ossbauer spectral analysis. Therefore it is not possible to rule out the presence of an incommensurate spin density wave on the basis of the $^{75}$As NMR data.
\end{abstract}

\submitto{\NJP}
\maketitle

\section{Introduction}
The nature of magnetism in the recently discovered iron pnictides superconductors \cite{kami} is an intensively studied topic, due to its possible coexistence with superconductivity and relation to the pairing mechanism \cite{Mazin09}. In the so-called ``122 family'', the parent compounds AFe$_2$As$_2$, where A=Ba, Ca, Sr or Eu, present a transition to a spin density wave (SDW) phase which is commensurate with the lattice spacing. When substituting Fe with another metal, like Co or Ni (the dopant), the N\'eel temperature gradually decreases and superconductivity eventually emerges at higher dopant concentrations. In Ba(Fe$_{1-x}$Co$_x$)$_2$As$_2$, superconductivity and magnetism coexist locally below the critical temperature $T_c$ \cite{Laplace09,Julien09}, i.e. there occurs a so-called mixed phase with spatial coexistence of the two phases. By contrast, in some pnictides belonging to another family, e.g. LaFeAsO$_{1-x}$F$_x$ \cite{Luetkens09}, magnetism vanishes when superconductivity sets in. Magnetic order in all AFe$_2$As$_2$ is antiferromagnetic (AF) and the spin arrangement is formed by AF stripes along $\boldsymbol{a}$ and ferromagnetic stripes along $\boldsymbol{b}$, in the orthorhombic notation \cite{Lumsden10}. The Fe ordered moment is small, of the order of 1\,$\mu_B$, which, together with the metallic behavior in the normal state, suggests the presence of itinerant magnetism. This is believed to be induced by a Fermi surface instability driven by the nesting between an electron and a hole pocket \cite{Singh09-1,Dong08}, as seen by angle-resolved photoemission studies (ARPES) in BaFe$_2$As$_2$ \cite{Fink09}, as well as in electron-doped Ba(Fe$_{1-x}$Co$_x$)$_2$As$_2$ \cite{Brouet09} and hole-doped Ba$_{1-x}$K$_x$Fe$_2$As$_2$ \cite{Ding08,Liu08}.

In doped systems, it was suggested that an imperfect nesting between the hole and electron pockets should lead to the formation of an incommensurate spin density wave (IC-SDW) below the magnetic transition temperature \cite{Park10}, as occurs in Cr \cite{Rice70}. Several theoretical studies have shown that coexistence of superconductivity and magnetism is favoured by the presence of an IC-SDW \cite{Vorontsov09,Vorontsov10,Cvetkovic09-PRB}. Therefore, the observation of an IC-SDW is an important issue to verify the various theories put forward in order to describe the physics of pnictide superconductors.
%From the experimental point of view, an IC-SDW was clearly observed by neutron scattering in chalcogenides Fe$_{1+y}$Se$_x$Te$_{1-x}$~\cite{Bao09,Wen09,Khasanov09}, but this issue is not yet settled in pnictides. 
Experimentally, a phase transition from commensurate to incommensurate SDW was recently reported in SrFe$_2$As$_2$ under pressure \cite{Kitagawa09}, based on NMR measurements. In  Ba(Fe$_{1-x}$Co$_x$)$_2$As$_2$, the presence of an IC-SDW was inferred from local techniques such as NMR \cite{Laplace09}, M\"ossbauer spectroscopy \cite{Bonville10} and $\mu$SR \cite{Marsik10}, but was not observed by neutron scattering \cite{Christianson09,Pratt09} until recently, and only in a narrow concentration range close to a Co doping level $x$=0.055 \cite{Pratt11}. Therefore, it was claimed that the perturbation of magnetic order induced by the dopant is responsible for the effects observed in Refs.\cite{Laplace09,Bonville10}. In Ba(Fe$_{1-x}$Ni$_x$)$_2$As$_2$, $^{75}$As NMR lineshapes were interpreted in the framework of an empirical model according to which the magnetic order is commensurate, and the dopant induces perturbations of the Fe moment amplitude \cite{Dioguardi10, Dioguardi11-condmat}. In Ba(Fe$_{1-x}$Rh$_x$)$_2$As$_2$, M\"ossbauer spectra were interpreted assuming either an IC-SDW or the presence of disorder \cite{Wang11}, while neutron scattering did not detect any incommensurability in this family \cite{Kreyssig10}.

In this paper, we present M\"ossbauer results obtained on a powder sample of BaFe$_2$As$_2$ and in single crystal mosaics of Ba(Fe$_{1-x}$Ni$_x$)$_2$As$_2$ with $x$=0.01 and 0.03. In the doped systems, we show that a distribution of Fe hyperfine fields is present in the magnetic phase, similar to what was observed in Ba(Fe$_{1-x}$Co$_x$)$_2$As$_2$ \cite{Bonville10}. We interpret the M\"ossbauer data using the two models for an incommensurate spin density wave and for a dopant-induced Fe moment perturbation. Then, we present simulations of NMR spectra from Refs.\cite{Dioguardi10,Dioguardi11-condmat} assuming an IC-SDW modulation with the parameters obtained from the fit of the M\"ossbauer spectra. This is motivated by the observation made in Ref.\cite{Dioguardi10} that the NMR data rule out the presence of IC-SDW in Ni doped compounds.

\section{Experimental and sample characterization}

BaFe$_2$As$_2$ in powder form was prepared by solid-state reaction and it presents a magnetic transition at $T_{\rm{SDW}}\sim139.5$\,K \cite{Rullier-Albenque09}. Single crystals of Ba(Fe$_{1-x}$Ni$_x$)$_2$As$_2$ with $x=$0.01 and 0.03 were grown by the self-flux method, as described in detail in Ref.\cite{Olariu11}. The system is not superconducting at $x=0.01$, while at $x=0.03$ superconductivity occurs below $T_c \simeq11$\,K. Susceptibility measurements on a single crystal show that the whole sample is superconducting below $T_c$. Resistivity measurements performed on crystals from the same batches showed them to be very homogenous \cite{Olariu11}, with a magnetic transition towards a spin density wave at $T_{\rm{SDW}}\simeq 112.5$\,K for $x=0.01$ and $\simeq 47$\,K for $x=0.03$. Ni-doped platelet-like crystals were glued in mosaic on aluminium scotch tape with the ($a,b$) plane perpendicular to the $\gamma$ beam. The scotch tape contains iron impurities in very small amounts, which give rise to a quadrupolar doublet with 0.07\% resonant absorption. Because the resonant absorption due to the BaFe$_2$As$_2$ phase is around 2\%, the iron impurities give a negligible contribution to the spectra. Spectra were recorded using a commercial $^{57}$Co$^*$:Rh $\gamma$-ray source mounted on an electromagnetic drive with linear velocity signal. The width calibration of the source was done with a thin K$_4$Fe(CN)$_6$. 3H$_2$O absorber and the half width at half maximum (HWHM) is found to be 0.128\,mm/s. The velocity and linearity calibration were performed with an enriched $\alpha$-Fe$_2$O$_3$ absorber and a commercial Wissel laser interferometer. All spectra are shown in a velocity range of $\pm$2\,mm/s. However, the spectra in pure BaFe$_2$As$_2$ and the 4.2\,K spectrum in Ba(Fe$_{0.99}$Ni$_{0.01}$)$_2$As$_2$ were recorded in a velocity range of $\pm$3\,mm/s.

\section{$\textrm{BaFe}_2\textrm{As}_2$}

\begin{figure}
\begin{center}
\includegraphics[width=7 cm]{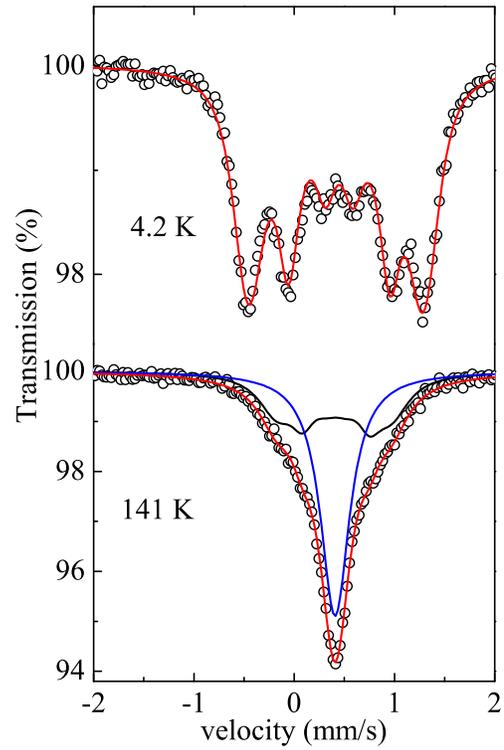}
\caption{M\"ossbauer absorption spectra on $^{57}$Fe in BaFe$_2$As$_2$ in the SDW phase (4.2\,K) and close to the SDW to paramagnetic transition (141\,K). Lines are fits as described in the text.}
\label{Pur_spectres}
\end{center}
\end{figure}

\begin{figure}
\begin{center}
\includegraphics[width=8 cm]{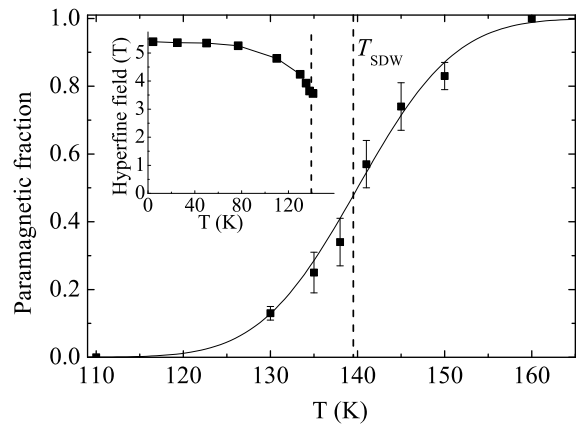}
\caption{Temperature dependence of the paramagnetic fraction in BaFe$_2$As$_2$ around $T_{\rm{SDW}}$, extracted from the fit of the M\"ossbauer spectra. The vertical dashed lines indicate $T_{\rm{SDW}} \simeq$139.5\,K. Inset: temperature dependence of the hyperfine field. The lines are guides for the eye.}
\label{Pur_ParamagFraction}
\end{center}
\end{figure}

In the high temperature paramagnetic range, the M\"ossbauer spectra are well fitted by a single Lorentzian-shaped line of HWHM 0.15\,mm/s. In the magnetically ordered phase below $T_{\rm{SDW}}\simeq 139.5$\,K (Fig.\ref{Pur_spectres}) and up to about 125\,K, spectra are well fitted to a 6-line pattern due to a single hyperfine field of 5.4\,T at 4.2\,K. This is expected for a commensurate magnetic order and it is in good agreement with similar measurements presented in Refs.\cite{Rotter09,Rotter08-PRB,Nowik10}, as well as with neutron scattering \cite{Huang08} and NMR \cite{Kitagawa08} results. A peculiarity of these hyperfine field spectra is that the HWHM of the middle lines, 0.134(4)\,mm/s, is equal within uncertainties to that of the inner lines, while the HWHM of the outer lines is 0.168(4)\,mm/s, significantly larger. This unusual spectral effect is observed in other published spectra at 4.2\,K \cite{Rotter09,Rotter08-PRB} and it cannot be due to a mere distribution of hyperfine fields, which would yield increasing broadenings for the inner, middle and outer lines. It can be accounted for by correlated distributions of quadrupolar interaction (QI) and isomer shift (IS). The presence of a  QI distribution at the $^{57}$Fe site is in line with the $^{75}$As NMR spectra in BaFe$_2$As$_2$, which indicate a sizeable distribution of quadrupolar frequencies at the As site \cite{Kitagawa08}. Correlated distributions of QI and IS have been observed in amorphous or glassy materials \cite{Dunlap98,Verma85}, but they are not expected in an ordered metallic material such as BaFe$_2$As$_2$. We note that they can be inferred to occur in charge density wave (CDW) phases, but such a phase has not been reported in the iron pnictides.

The thermal variation of the hyperfine field is shown in the inset of Fig.\ref{Pur_ParamagFraction}. Its weak temperature dependence suggests a transition with first order character. In the temperature range from 130 to 150\,K close to $T_{\rm SDW}$, the spectra cannot be fitted to a single sextet. An additional single-line paramagnetic component develops at the expense of the magnetically ordered one as temperature increases above 120\,K. The fit with two components obtained at 141\,K is shown in Fig.\ref{Pur_spectres}. The presence of this relatively narrow magnetic/paramagnetic coexistence region (see Fig.\ref{Pur_ParamagFraction}) is another indication of the first order character of the transition. This is in good agreement with the $^{75}$As NMR data in BaFe$_2$As$_2$ \cite{Kitagawa08}. In BaFe$_2$As$_2$, the paramagnetic fraction develops from 0 to 100\% over a 15\,K range  on each side of the transition ($\pm$0.1$T_{\rm{SDW}}$). This is quite different from the behaviour in Co-doped systems, where  coexistence is observed to occur on a much wider temperature range \cite{Bonville10}, up to about 1.5 $T_{\rm{SDW}}$ above the magnetic transition.

\section{$\textrm{Ba(Fe}_{1-x}\textrm{Ni}_x)_2\textrm{As}_2$}

\subsection{Interpretation of M\"ossbauer spectra}

\begin{figure}
\begin{center}
\includegraphics[width=11 cm]{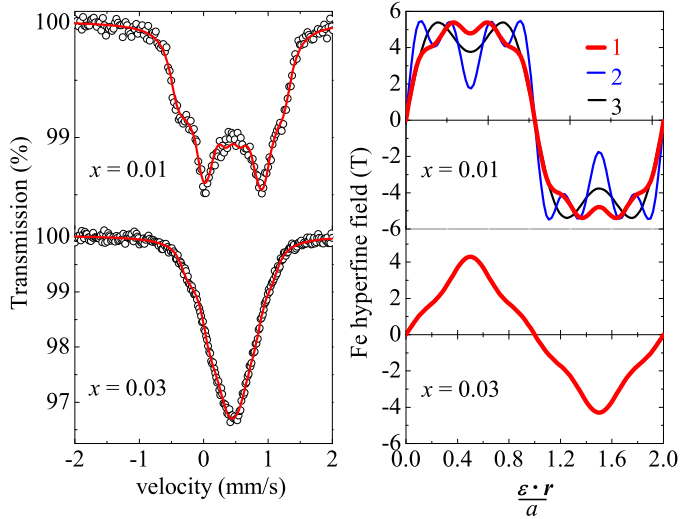}
\caption{{\bf Left:} M\"ossbauer spectra in Ba(Fe$_{1-x}$Ni$_x$)$_2$As$_2$ obtained at 4.2\,K for $x$=0.01 and 0.03. Continuous lines are fits assuming an incommensurate modulation of Fe moments. {\bf Right:} modulations that fit well the spectra shown in the left panel, with parameters from Table I. The thick red lines correspond to the fits shown on the left.}
\label{Ni_spectres4K}
\end{center}
\end{figure}

Mosaic M\"ossbauer spectra obtained for Ba(Fe$_{1-x}$Ni$_x$)$_2$As$_2$ with $x$=0.01 and 0.03 with the $\gamma$-beam along the {\bf \emph{c}} axis are shown in Fig.\ref{Ni_spectres4K} for $T=4.2$\,K. These spectra do not display the six narrow line pattern expected for a single magnetic environment, as in the undoped compound. This indicates a distribution of Fe hyperfine fields, similar to what was observed in Ba(Fe$_{1-x}$Co$_x$)$_2$As$_2$ \cite{Bonville10}. Our data is in agreement with that reported in Refs.\cite{Nowik10, Felner11} for the Ni content $x$=0.016 and 0.024. In these works, the spectra were fitted to a superposition of commensurate and sine-wave incommensurate subspectra \cite{Nowik09}. Here, we fit the spectra assuming either an incommensurate spin density wave with several harmonics, or a dopant-induced perturbation model. We focus mostly on the $x=0.01$ Ni concentration in order to take advantage of the good resolution of the spectra. For $x \geq 0.016$, due to the decrease of the Fe magnetic moment, spectra become almost featureless and therefore lack the resolution necessary for our analysis.

Spectra can be very well fitted by assuming an incommensurate modulation of the Fe moments with several harmonics (Fig.\ref{Ni_spectres4K}, left panel) as described in Ref.\cite{Bonville10}. According to this model, the hyperfine field $H_{hf}$ can be developed as a Fourier series along the propagation vector $\boldsymbol \epsilon$ of the incommensurate modulation:
\begin{equation}
H_{hf}(\boldsymbol{\epsilon \cdot r}) =\sum_{k=0}^{n}{h_{2k+1}\rm{sin}}[(2k+1)\frac{\pi}{a}\boldsymbol{\epsilon \cdot r}].
\label{hhf}
\end{equation}
The fit of the M\"ossbauer spectra allows us to extract the parameters $h_{2k+1}$. Due to the local character of the technique,  the vector $\boldsymbol{\epsilon}$ cannot be determined. Fourier coefficients obtained for $x$=0.01 and 0.03 are given in Table I and the corresponding modulations are shown in Fig.~\ref{Ni_spectres4K} (right panel). Note that there are several sets of parameters that yield equally good fits of the spectra. The shape of the modulation appears to evolve with doping: for $x=0.01$ it is close to a "square-wave" while it approaches a sine-wave for $x=0.03$, similar to Ba(Fe$_{1-x}$Co$_x$)$_2$As$_2$. The relative intensities of the lines are close to 3:4:1:1:4:3, as expected when the hyperfine field, and thus the Fe magnetic moment, is perpendicular to the $\gamma$-beam.  This is in good agreement with neutron scattering measurements in Ba(Fe$_{1-x}$Ni$_x$)$_2$As$_2$ at low doping levels, showing that moments are parallel to {\bf \emph{a}} as in pure BaFe$_2$As$_2$.

\begin{table}
\caption{Fourier coefficients (in T) obtained for Ni-doping with $x=0.01$ (three sets yielding equally good fits) and $x=0.03$ (only one set is shown, although several sets fit the data equally well).}
\label{coeff}
\begin{center}
\begin{tabular}{|c|c|c|c|c|c|c|c|c||} \hline
$x$   & N$^{\rm{o}}$ & $h_1$ & $h_3$ & $h_5$ & $h_7$  \\ \hline
0.01 & 1 & 5.95 & 1.04 & 0.312 & 0.444 \\ \hline
0.01 & 2 & 5.30 & 2.60 & 0.486 & 1.45 \\ \hline
0.01 & 3 & 5.69 & 1.93 & 0 & 0 \\ \hline
0.03  & - & 3.61 & -0.43 & 0.26 & 0  \\ \hline
\end{tabular}
\end{center}
\end{table}

So far we have shown that the hypothesis of IC-SDW leads to very good fits of the M\"ossbauer spectra. We now show that good fits can also be obtained with the empirical model described in Ref.\cite{Dioguardi10}, based on first-principle calculations \cite{Kemper09}, assuming that the dopant induces a perturbation of the Fe moment amplitude. Within this model, the Fe moment value $m$(\emph{\textbf{r}}) at distance \emph{\textbf{r}} from the origin, is given by the expression:
\begin{equation}
m(\boldsymbol{r})=m_0 \Phi_x(\boldsymbol{r})\ \textrm{cos}(\boldsymbol{Q_0\cdot r}), \label{mdop}
\end{equation}
where $\Phi_x=1-\alpha \sum_i$ exp[-$(\boldsymbol{r-r_i})^2/\lambda^2$]. In this latter expression, $\boldsymbol{r_i}$ is the distance to the $i$th dopant atom, the sum runs over all dopant atoms and $\boldsymbol{Q}_0=\frac{\pi}{a}(1,0,1)$ is the SDW vector for the commensurate pattern observed in BaFe$_2$As$_2$. The parameter $\alpha$ gives the reduction of the spin of the dopant atoms, while $\lambda$ describes the spatial extent of the perturbation and $m_0$ is an unperturbed Fe moment. This model allows one to reproduce correctly the $^{75}$As NMR spectra for different Ni concentrations \cite{Dioguardi10}, with parameters $\alpha=0.6$ and $\lambda=2.5a$. In the following, we shall refer to this model as the ``perturbation'' model. 

\begin{figure}
\begin{center}
\includegraphics[width=11 cm]{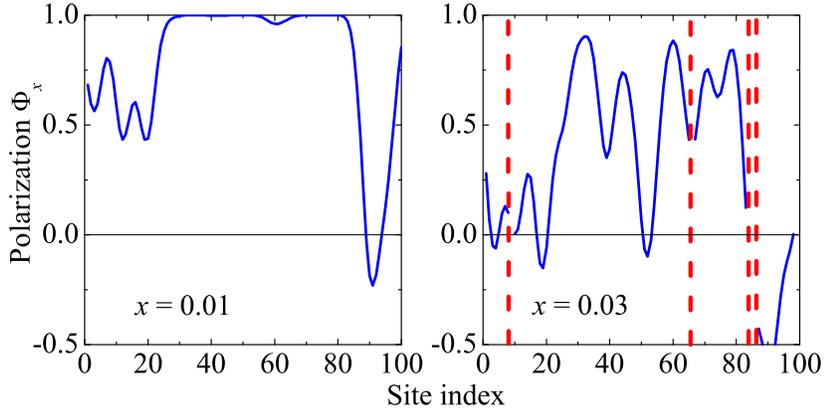}
\caption{Fe spin polarization calculated, for $x$=0.01 and 0.03, in the framework of the ``perturbation'' model along a line in the ($\boldsymbol a$,$\boldsymbol b$) plane parallel to $\boldsymbol a$. We use expression (\ref{mdop}) with parameters $\alpha$=0.6 and $\lambda=2.5a$. The red dashed lines mark the positions of the dopant atoms. According to this model, the periodicity of the magnetic structure should be completely broken for $x$=0.03, in disagreement with neutron scattering results.}
\label{Polarization}
\end{center}
\end{figure}

\begin{figure}
\begin{center}
\includegraphics[width=11 cm]{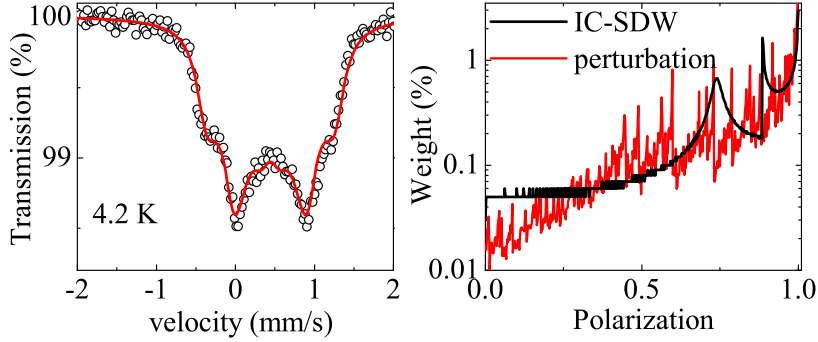}
\caption{{\bf Left:} M\"ossbauer spectrum at 4.2\,K for Ba(Fe$_{1-x}$Ni$_x$)$_2$As$_2$ with $x=0.01$, fitted to the ``perturbation'' model. {\bf Right:} histogram of the reduced Fe polarizations for the IC-SDW and ``perturbation'' models for $x$=0.01.}
\label{Ni_spectre4Khist}
\end{center}
\end{figure}

In order to visualize the effect of the dopant atom on its magnetic environment within this model, we chose a (300 $\times$ 300) point mesh representing one Fe-As ($\boldsymbol a$,$\boldsymbol b$) plane and placed Ni atoms at random in this plane according to the concentration. We compute the Fe polarization along a given direction as a function of the distance from the origin for $x$=0.01 and 0.03 (Fig.\ref{Polarization}). For $x=0.01$, the polarization remains close to 1 for most sites, although destructive interferences lead to a few Fe sites with strongly depleted or zero moments. For $x=0.03$, this model leads to an average polarization much smaller than 1 and even negative for a few sites. The disorder induced by the presence of Ni dopants would be so strong that the periodicity of the magnetic structure would be completely broken. If this were the case, no magnetic order could have been detected by neutron scattering. However, Bragg peaks are clearly observed at concentrations close to $x=0.03$ \cite{Wang10}, which shows that the model is  not valid beyond the very dilute limit. Note that for even larger dopant content, the polarization of all sites would turn negative and smaller than -1, which is unphysical. We therefore restricted our simulations of the M\"ossbauer spectra to $x=0.01$. Good agreement with the data was found for this concentration for a wide range of parameter values, with $\alpha$ varying from 0.4 to 0.8 and $\lambda$ from $2a$ to $3a$. The obtained fit is shown in the left panel of Fig.\ref{Ni_spectre4Khist} for $\alpha=0.6$ and $\lambda=2.5a$. Note that even for $x=0.01$, the rather large values of $\alpha$ and $\lambda$ seem to overestimate the influence of magnetic disorder with respect to what is predicted by theoretical calculations \cite{Kemper09}.

One can wonder why both the IC-SDW and ``perturbation'' models reproduce the M\"ossbauer spectra well. This is in fact due to similar features of the distributions of Fe hyperfine fields for both models, as shown in the right panel of Fig.\ref{Ni_spectre4Khist}. Both histograms are very asymmetric, with a quasi-divergence at the maximum field value and a continuous distribution extending down to zero polarisation. For the ``perturbation'' model, the noise is due to the random lattice positions of the dopant atoms.

\begin{figure}
\begin{center}
\includegraphics[width=8 cm]{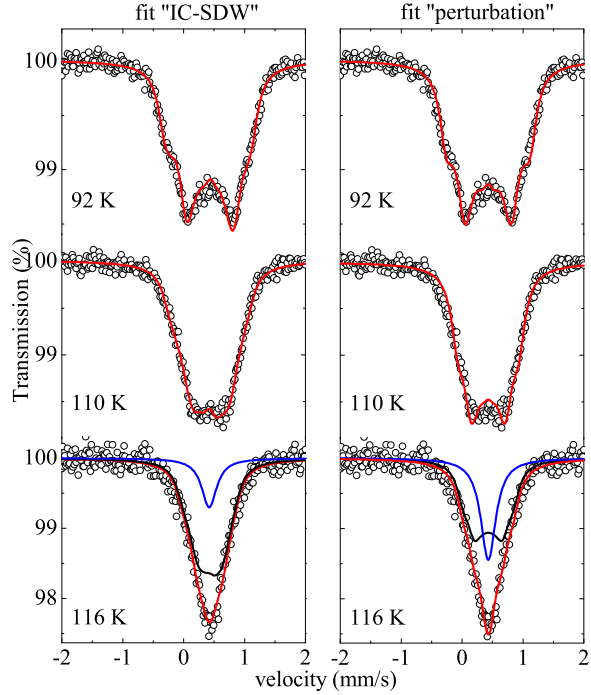}
\caption{M\"ossbauer spectra at different temperatures for Ba(Fe$_{1-x}$Ni$_x$)$_2$As$_2$ with $x=0.01$. Continuous lines are fits assuming either an incommensurate modulation (left panel) or an dopant-induced distribution of the Fe hyperfine field (right panel). At 116\,K, a Lorentzian-shaped component was added to the fit in order to represent the paramagnetic fraction of the sample.}
\label{Ni1p_spectres}
\end{center}
\end{figure}

\begin{figure}
\begin{center}
\includegraphics[width=8 cm]{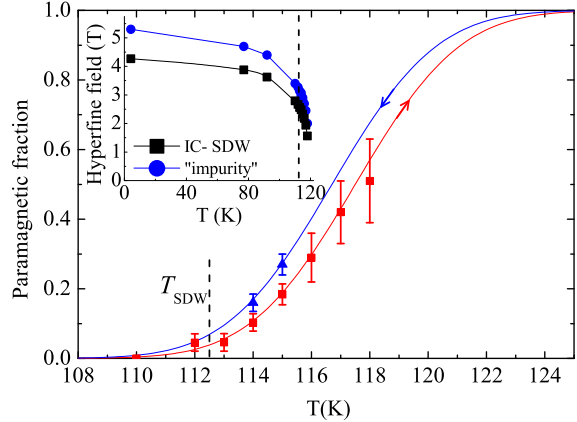}
\caption{Ba(Fe$_{1-x}$Ni$_x$)$_2$As$_2$ with $x=0.01$: temperature dependence of the paramagnetic fraction for increasing and decreasing temperature, derived from the fit of M\"ossbauer spectra with an IC-SDW.  Inset: temperature dependence of the hyperfine field, obtained from both fits with an IC-SDW and dopant-induced perturbation. The vertical dashed lines indicate $T_{\rm{SDW}} \simeq 112.5$\,K. Lines are guides for the eye.}
\label{Ni1p_ParamagFraction}
\end{center}
\end{figure}

Spectra recorded at higher temperatures for $x=0.01$ are shown in Fig.\ref{Ni1p_spectres}. All the spectra measured below $T_{\rm{SDW}}\simeq 112.5$\,K can be well fitted either with the IC-SDW or the ``perturbation'' model with the parameters $\alpha=0.6$ and $\lambda=2.5a$, as described above. The IC-SDW model yields systematically better fits with $\chi^2 \sim$1.10 - 1.25, while fits with the ``perturbation'' model have $\chi^2 \sim$1.5 - 2. However, this is not very significant because the IC-SDW model involves more parameters. Fits with the IC-SDW model yield similar shapes of the modulations in the temperature range from 4.2 to 110\,K. Above $T_{\rm{SDW}}$ determined by resistivity measurements, a paramagnetic component develops at the expense of the magnetic one in a temperature range of 0.1$T_{\rm{SDW}}$, similar to pure BaFe$_2$As$_2$, indicative of a first order transition. The paramagnetic fraction, shown as a function of temperature in Fig.\ref{Ni1p_ParamagFraction}, grows rapidly with $T$, so that at $T\sim120$\,K the sample is almost fully paramagnetic. Note that when the fraction of the magnetic phase is less than $\sim$50\%, it is impossible to perform an accurate quantitative determination of the paramagnetic fraction, due to the lack of spectral resolution. Therefore, no experimental data are shown for temperatures above 118\,K. In the temperature range from 110 to 116\,K, fits with the IC-SDW model were performed by assuming that the shape of the modulation does not evolve with $T$, in order to limit the number of fitting parameters. Hysteresis is observed between warming and cooling procedures. This, together with the steep decrease of the hyperfine field when approaching $T_{\rm{SDW}}$ (Fig.\ref{Ni1p_ParamagFraction} inset), confirms the first order nature of the transition. Here the hyperfine field $H_{\rm{hyp}}$ was derived from the fit with both the ``perturbation'' and IC-SDW models. In the former case, $H_{\rm{hyp}}$ is a free parameter proportional to $m_0$. In the latter case, $H_{\rm{hyp}}$ is the average over a period of the incommensurate modulation, given by the expression:
\begin{equation}
H_{\rm{hyp}}=(\frac{1}{2}\sum_k h_{2k+1}^2)^{0.5}
\label{hhk}
\end{equation}
We also fitted the spectrum recorded at 4.2\,K in Ba(Fe$_{1-x}$Co$_x$)$_2$As$_2$ \cite{Bonville10} for $x$=0.014 to the ``perturbation'' model. A rather good fit is obtained (not shown) with parameters $\alpha\simeq 0.37$ and $\lambda\simeq 2.5a$. The value of $\lambda$ which describes the extent of the spatial distribution is found to be the same as for Ni doping.

\subsection{Simulations of $^{75}$As NMR spectra assuming an IC-SDW}

We have shown in the previous section that, in BaFe$_2$As$_2$ doped with Co or Ni, the $^{57}$Fe M\"ossbauer spectra can be equally well reproduced assuming either an IC-SDW or a dopant-induced perturbation model. So the M\"ossbauer spectra do not allow us to distinguish between the two. Recently, it was claimed that the IC-SDW assumption yielded poor fits to the $^{75}$As NMR spectra in Ni doped BaFe$_2$As$_2$ \cite{Dioguardi10,Dioguardi11-condmat}, and that this fact favoured the ``perturbation'' model. In these latter works, the simulations of the NMR spectra with the IC-SDW assumption were carried out using the incommensurate modulation inferred from the NMR data at $x$=0.06 Co-doping \cite{Laplace09}. Here, we perform simulations of the NMR spectra with the incommensurate modulations derived from the M\"ossbauer spectra and we show that good agreement is achieved.

We consider a plane with (500 $\times$ 500) Fe sites and calculate, for each As site, the corresponding hyperfine field. The $^{75}$As hyperfine Hamiltonian can be written as:
\begin{equation}
{\cal H}_{hf}=\gamma \hbar \hat{\textbf{I}} \cdot \sum_i \mathbb{B}_i \cdot \boldsymbol{m(r_i)}.
\label{hfrmn}
\end{equation}
Here the sum runs over the four Fe nearest neighbors of an As nucleus with moment $\boldsymbol{m}$, $\gamma$ is the gyromagnetic ratio of $^{75}$As ($\gamma=7.292$\,MHz/T), $\hat{\textbf{I}}$ is the nuclear spin operator and $\mathbb{B}$ the hyperfine tensor. Four hyperfine tensor components are known: $B_{aa}=B_{bb}=0.88$\,T/$\mu_B$, $B_{cc}=0.47$\,T/$\mu_B$ and $B_{ac}=0.43$\,T/$\mu_B$, determined previously in pure BaFe$_2$As$_2$ \cite{Kitagawa08}. The fifth component $B_{ab}$ is unknown, but it has to be taken into account when the external field is perpendicular to the $\boldsymbol c$ axis \footnote{However, if the propagation vector $\boldsymbol\epsilon$ of the incommensurate modulation is parallel to the $\boldsymbol a$ axis, the term in the Hamiltonian involving $B_{ab}$ is zero.}. The Fe magnetic moment at position $\boldsymbol{r}$ is given by:
\begin{equation}
m(\boldsymbol{r})=m_0 \ \textrm{cos}(\boldsymbol{Q_0 \cdot r})\ \sum_{k=0}^{n} \frac{h_{2k+1}}{h_1} \ \textrm{sin}[(2k+1)\frac{\pi}{a} \boldsymbol{\epsilon \cdot r}],
\label{mhk}
\end{equation}
the total propagation vector being $\boldsymbol{\tau}=\boldsymbol{Q}_0+\frac{\pi}{a}\boldsymbol{\epsilon}$. The parameters $h_{2k+1}$ are known from the M\"ossbauer spectra, therefore the free parameters are $\boldsymbol{\epsilon}$, $m_0$ and $B_{ab}$.

\begin{figure}
\begin{center}
\includegraphics[width=9 cm]{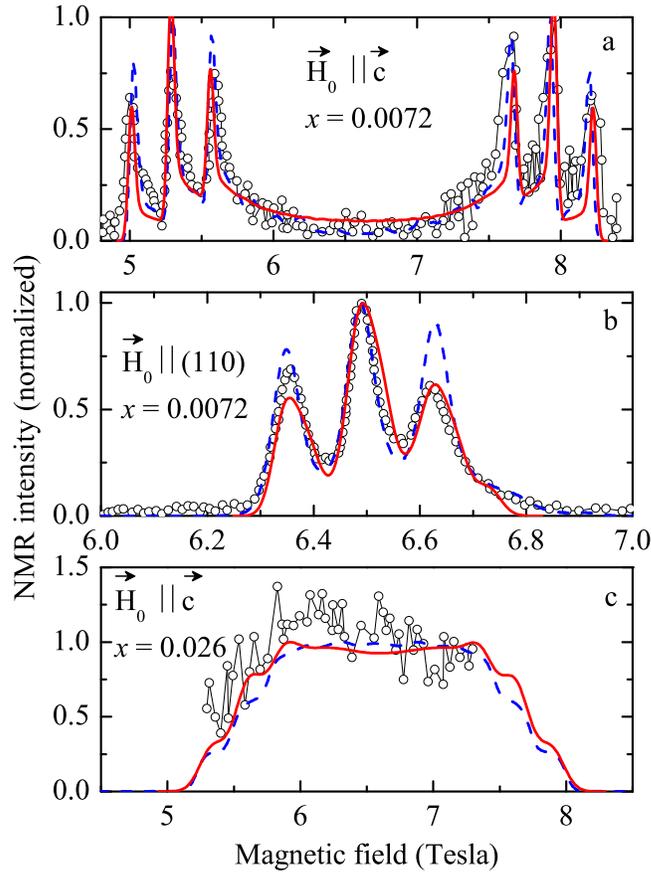}
\caption{$^{75}$As NMR spectra in Ba(Fe$_{1-x}$Ni$_x$)$_2$As$_2$ taken from Refs.\cite{Dioguardi10, Dioguardi11-condmat}. For $x$=0.0072: {\bf a}) the external magnetic field applied parallel to the $\boldsymbol c$ axis, and {\bf b}) along the [110] direction. For $x=0.026$: {\bf c}) the external magnetic field applied parallel to the $\boldsymbol c$ axis. The solid lines are simulations with the IC-SDW model and the dashed lines are simulations with the ``perturbation'' model.}
\label{RMN-Moss}
\end{center}
\end{figure}

In Fig.\ref{RMN-Moss} we show the NMR spectra from Refs.\cite{Dioguardi10,Dioguardi11-condmat} for two samples in the low doping regime, with the external field parallel to the $\boldsymbol c$ axis or along [110]. Simulations with the dopant-induced perturbation model are shown as dashed lines, with parameters $\alpha=0.6$ and $\lambda=2.5a$. The quadrupolar frequency was taken as $\nu_Q=2.0$\,MHz and the magnetic moments were 0.77\,$\mu_B$ for $x=0.0072$ and 0.65\,$\mu_B$ for $x=0.026$.

For the simulations of the NMR spectra for $x=0.0072$ assuming IC-SDW, we considered the modulation corresponding to the $x$=0.01 M\"ossbauer spectrum. The best fits to the NMR spectra were found with the set of parameters n$^{\rm{o}}$ 1 in Table I. Regardless of the propagation vector direction ($\boldsymbol{a}$, $\boldsymbol{b}$ or [110]), we find very good agreement between simulations and the data for realistic values of $\epsilon$, $m_0$ and $B_{ab}$. Continuous lines in Figs.\ref{RMN-Moss} {\bf a} and {\bf b} show the simulations obtained when the modulation is along [110], with the external field parallel or perpendicular to the $\boldsymbol c$ axis. The parameters are $\epsilon \simeq 0.13$, $m_0=0.91$\,$\mu_B$ and $B_{ab}=0.6$\,T/$\mu_B$. The obtained distribution was convoluted with a Gaussian with full width at half maximum (FWHM) 0.05\,T, in order to take into account the disorder induced by the Ni substitution. Simulations are not very sensitive to the value of the incommensurate parameter if it is kept small. Good fits are obtained in the range $\epsilon \leq 0.2$.

In a similar manner, we simulated the NMR spectrum obtained for $x=0.026$ assuming IC-SDW with the Fourier components for $x=0.03$ in Table I and with the same $\epsilon$ and $B_{ab}$ values as above. Figure \ref{RMN-Moss} c shows the simulation obtained for a modulation along [110], with $m_0=0.55$\,$\mu_B$, convoluted with a Gaussian of FWHM 0.2\,T. It is quite close to the experimental lineshape.

\section{Discussion and Conclusion}

Our $^{57}$Fe M\"ossbauer spectroscopy measurements on the pure and Ni-doped BaFe$_2$As$_2$ indicate the presence of a first order transition towards a spin density wave state. In pure BaFe$_2$As$_2$, the SDW is commensurate, in agreement with neutron scattering results, and the first order nature of the transition is in line with $^{75}$As NMR results.

In Ba(Fe$_{1-x}$Ni$_x$)$_2$As$_2$ with $x=0.01$ and $x=0.03$, the magnetic phase is characterized by a broad distribution of Fe hyperfine fields. We analyzed our results in relation to NMR measurements in the framework of two different models: an incommensurate spin density wave and a dopant-induced perturbation of the Fe moments. We have shown that both $^{75}$As NMR and M\"ossbauer spectra can be  fitted well with either model, contrary to what is stated in Refs.\cite{Dioguardi10,Dioguardi11-condmat}. Therefore, at low doping it is not possible to conclude from the available NMR or M\"ossbauer data whether the ``perturbation'' or the IC-SDW model is relevant in Ni doped BaFe$_2$As$_2$. At larger doping, a more elaborate model is necessary in order to track the role of disorder, because the ``perturbation'' model collapses even for moderate concentrations such as $x=0.03$.

In the more studied Ba(Fe$_{1-x}$Co$_x$)$_2$As$_2$ family, the presence of IC-SDW is still a matter of debate. While IC-SDW was inferred from local techniques, resonant x-ray scattering measurements for $x=0.047$ did not find any sign of incommensurability along $\boldsymbol{a}$ or $\boldsymbol{b}$, down to relatively small values of $\epsilon$ \cite{Kim10-X-rayResonant}. Neutron scattering detected an IC-SDW only recently, in a small concentration range above $x=0.056$, with a propagation vector $\boldsymbol\epsilon$ along $\boldsymbol{b}$ and $\epsilon$ values ranging from 0.02 to 0.03 \cite{Pratt11}. The M\"ossbauer spectrum for this concentration is well fitted with these parameters \footnote{unpublished data}. One can wonder whether IC-SDW could be present at all concentrations in the low doping regime of Ba(Fe$_{1-x}$Co$_x$)As$_2$, because the direction of the propagation vector of the modulation could change with $x$, following variations in the Fermi surface nesting. This would further complicate the detection of IC-SDW by neutron scattering and resonant x-ray measurements.

While disorder due to substitutions is inherent in Ba(Fe$_{1-x}$Ni$_x$)$_2$As$_2$ and Ba(Fe$_{1-x}$Co$_x$)$_2$As$_2$, there are a few stoichiometric systems in which the signature of incommensurability was observed and the presence of disorder can be ruled out. In the ``122 family'', the stoichiometric system SrFe$_2$As$_2$ presents, under pressure, a crossover versus temperature from commensurate to incommensurate AF, seen by $^{75}$As NMR measurements \cite{Kitagawa09}. In NaFeAs, a similar crossover was evidenced at ambient pressure by $^{23}$Na and $^{75}$As NMR \cite{Kitagawa11,Klanjsek11}. This system presents filamentary superconductivity below 23\,K and, similar to other Fe-based pnictides, shows a structural phase transition from tetragonal to orthorhombic at 50\,K, followed by a magnetic one towards an antiferromagnetically ordered state at 40\,K. These observations suggests that incommensurability is not only related to doping but rather, might be connected to subtle changes of the Fermi surface. 
%It would be interesting to study this compound by M\"ossbauer spectroscopy and compare the spectral shape to that obtained for Ni and Co doping.

In conclusion, we have shown that at very low doping levels, both IC-SDW and dopant-induced perturbation models fit the NMR and M\"ossbauer spectra well in Ba(Fe$_{1-x}$Ni$_x$)$_2$As$_2$. However, the latter model collapses even for moderate dopant concentrations $x=0.03$, because it overestimates the disorder effect. Further theoretical calculations are required in order to sort out the precise influence of the dopant on the magnetic properties of pnictides. On the other hand, the IC-SDW model successfully reproduces the spectra over the whole doping region. This, together with the observation of IC-SDW in the stoichiometric systems mentioned above and the similarity with the Co family, suggests that IC-SDW might also be present in the Ni family.

\ack
Authors wish to thank Julien Bobroff and
Henri Alloul for helpful discussions and reading of the manuscript, and Sylvie Poissonnet (SRMP/CEA) for the chemical analysis of the samples. This work was supported by the ANR grant
PNICTIDES. A.O. acknowledges financial support from ``Triangle de la Physique''.
\\

%\bibliographystyle{iopart-num}
%\bibliography{E:/areta/Biblio/PNICTIDES}% Produces the bibliography via BibTeX.
%\input{E:/areta/Biblio/PNICTIDES.BIB}

\end{document}